\documentclass[conference]{IEEEtran}
\usepackage{cite}
\usepackage{amsmath}
\usepackage[papersize={8.5in,11in}, margin=0.75in]{geometry}
\usepackage{amsfonts}
\usepackage{graphicx}
\usepackage{subcaption}
\usepackage{algorithm}
\usepackage{algpseudocode}
\usepackage[font=small,skip=0pt]{caption}
\usepackage{booktabs}
\usepackage{pgfplots}
\usepackage{multirow}
\pgfplotsset{compat=newest}
\usepackage{array}
\begin{document}
\title{MER-SDN: Machine Learning Framework for Traffic Aware Energy Efficient Routing in SDN }

\author{\IEEEauthorblockN{Beakal Gizachew Assefa and Oznur Ozkasap}
\IEEEauthorblockA{Department of Computer Engineering\\
Koc University, Istanbul, Turkey}
\{bassefa13, oozkasap\}@ku.edu.tr}

\maketitle
\begin{abstract}
Software Defined Networking (SDN) achieves programmability of a network through separation of the control and data planes. It enables flexibility in network management and control. Energy efficiency is one of the challenging global problems which has both economic and environmental impact.  A massive amount of information is generated in the controller of an SDN based networks. Machine learning gives the ability to computers to progressively learn from data without having to write specific instructions. In this work, we propose MER-SDN: a machine learning framework for traffic aware energy efficient routing in SDN. Feature extraction, training, and testing are the three main stages of the learning machine. Experiments are conducted on  Mininet and POX controller using real-world network topology and dynamic traffic traces from SNDlib. Results show that our approach achieves more than 65\% feature size reduction, more than 70\% accuracy in parameter prediction of an energy efficient heuristics algorithm, also our prediction refine heuristics converges the predicted value to the optimal parameters values with up to 25X speedup as compared to the brute force method. 
\end{abstract}

\section{Introduction}

SDN is a widely accepted networking paradigm based on the concept of separation of control and data planes. It is implemented in a home network, campus networks, ISP, telecom, and cloud data centers. Major companies like Facebook, Yahoo, Microsoft, Huawei, Cisco, and Google has adopted SDN to their data centers and network equipment designs \cite{sdnsurvey,compsurvey}. 

Machine learning is used in various disciplines as a tool to discover a pattern in a structured, semi-structured and unstructured data. It has a global impact on the technology as it is useful in AI, genetics, computer vision, business prediction, and others. In SDN, because of the logically centralized controller, a massive amount of information is generated and stored instantly. With the ever-increasing network information, machine learning techniques are formidable and play a vital role in discovering knowledge from the stored network information \cite{sdnmlte2014,seer2016,2017SDNmlrouting}.

One of the most prominent challenges of the present world is energy since it has both economic and ecological issues. 10\% of the global energy consumption is due to ICT sector out of which 2\% is from network components. By 2020 the total electricity cost of cloud data centers is expected to increase by 63\% \cite{MAALOUL2017,stat}. SDN enables us to achieve traffic proportional energy consumption through dynamic re-routing of flows. The practical solution is to sleep/turn off underutilized components during low traffic load. However, there is a trade-off between performance and efficiency since turning off network components for sake of efficiency has an adverse effect on performance. 

In our previous work \cite{blackseacomassefa,blackseacomassefa2018}, we have proposed IP formulations and heuristics to maintain this trade-off. However, the efficiency of the three heuristics we proposed namely Next Shortest Path, Next Maximum Utility, and MEPT depend on the value of the utility interval parameters $Umin$ and $Umax$. In the previous work, we have determined these parameter values by brute force. 
   
In this work, we propose MER-SDN, a framework that combines the capabilities of SDN and machine learning for energy efficient routing. MER-SDN is implemented on the POX controller. It extracts the topology and traffic information and stores it in a repository. It also uses PCA (Principal Component Analysis) for feature size reduction and linear regression for training the model. 
MER-SDN is a generic framework that can be applied to a range of energy efficient approaches. In this work, however, we use it to predict the optimal values of the $Umin$ and $Umax$ parameters for the MEPT heuristics. We also propose a heuristics to maximize the accuracy of the predicted $Umin$ and $Umax$ values.

The contributions of this work are as follows.

\begin{itemize}

\item We propose a three module machine learning framework for traffic proportional energy saving in SDN. The modules are Traffic Manager, Topology Manager, and Learning Machine.

\item Most of the machine learning approaches used in SDN are for traffic classification, routing, intrusion detection, or attack prediction. To the best of our knowledge, we are the first in applying it to energy saving and performance combined.

\item We present a full-fledged machine learning method that starts from feature extraction, applies feature reduction, and also provides a heuristics to increase the accuracy of the predictor to 100\% in a constant time.

\item We present the cross-fold validation results that show how we choose the number of principal components for PCA for the training set. The results indicate more than 65\% feature size reduction.

\item Our model predicts $Umin$ and $Umax$ with an accuracy of more than 70\%. The refine heuristics we proposed to increase the accuracy converges to the optimal values with a speedup of 15 to 25 times as compared to the brute force approach.

\end{itemize}

The remainder of the paper is organized as follows. Section \ref{sec:related} presents related work. MER-SDN framework is discussed in section \ref{sec:mer-sdn}. Section \ref{sec:experiment} presents the experimental analysis of our approach. Conclusion and future work are discussed in section \ref{sec:conclusion}

\section{Related Work}\label{sec:related}

The use of machine learning techniques for energy efficiency in traditional networks has been studied \cite{trd2015}, where the techniques are applied in assisting resource management, power distribution, demand forecasting, workload prediction, virtual machine placement prediction, memory assignment, CPU frequency, and traffic classification. The techniques range from supervised learning, unsupervised learning, reinforcement learning, and hybrid combination.

A meta-layered machine learning approach composed of multiple modules is proposed \cite{sdnmlte2014}. The goal of this approach is to mimic the results of heuristics used in traffic engineering to maximize the quality of service (QoS). However, each neural network per module is trained separately and each trained model operates separately for each demand pair. The drawback of this approach is that it does not represent the relationships between the demands.

Seer is a configurable platform for network intelligence based on SDN,  Knowledge Centric Networking, and Big Data principles, where the goal is to accommodate the development of future algorithms and application that target network analytics \cite{seer2016}. It is also flexible in a sense that it allows high-level users to decide what network information to use for their goals. By focusing on reliability, the platform aspires to provide a scalable, fault-tolerant and real-time platform, of production quality.

Machine learning in SDN is also used in predicting the host to be attacked \cite{2016attack} using C4.5, Bayesian Network, Decision Table, and Naive-Bayes algorithms. Prediction of DDoS attack using neural network is implemented in NOX controller \cite{2014DDoSattack}.

Another machine learning based approach NeuRoute is a dynamic framework which learns a routing algorithm and imitates its results using neural networks in real-time.  NeuRoute is implemented on top of Google's TensorFlow machine learning framework and tested on POX controller. Experimental findings on  GEANT topology show that the NeuRoute is faster than dynamic routing algorithms \cite{2017SDNmlrouting}.

In contrast to the existing machine learning based solutions proposed for SDN, our framework models performance and energy efficiency at the same time. We present a method of representing traffic as features, perform feature size reduction using mathematically proven techniques, provide heuristics to increase the accuracy of the prediction to 100\%.  In our approach, we predict utility parameters $Umin$ and $Umax$ for the MEPT heuristics algorithm \cite{blackseacomassefa2018}.

\section{MER-SDN Framework Description}\label{sec:mer-sdn}

We propose MER-SDN framework that utilizes machine learning techniques to achieve traffic proportional energy efficiency in SDN. The objectives are to jointly formulate energy efficiency and network performance, to propose generalized heuristics algorithms, and to apply machine learning approaches on SDN controller that learn from traffic, network and solution history.

\begin{figure}[h]
    \vspace{-.4cm}
    \centering \includegraphics [width=0.9\linewidth,trim=4 4 4 4,clip]  {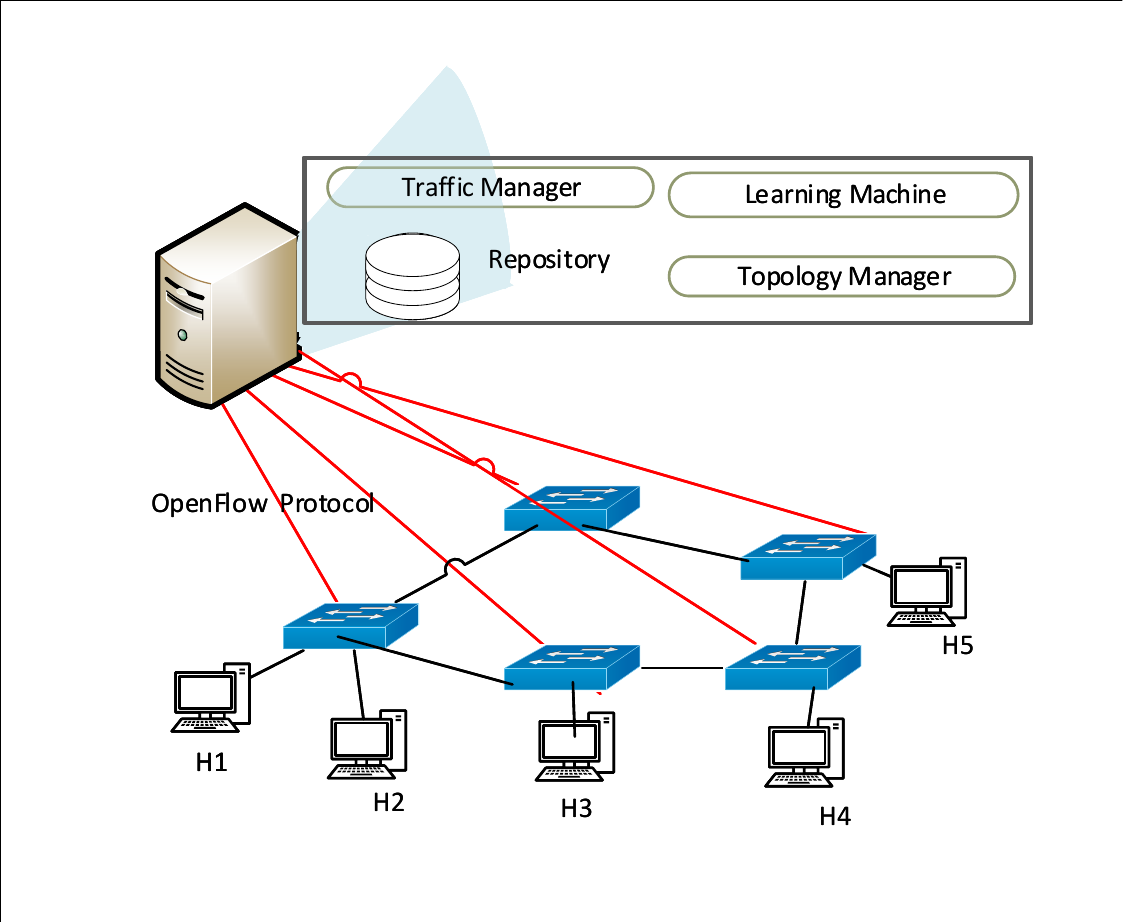}
    \caption{MER-SDN: Machine Learning Framework for Energy Efficient Routing in Software Defined Networking}
    \label{fig:taframework} 
\end{figure}

Figure \ref{fig:taframework} illustrates the MER-SDN framework consisting of three modules. The information of the traffic generated by the applications is passed to the traffic manager that stores details of traffic information in terms of source-destination pairs, rate, time a traffic demand arrives, and the total amount of flow. The status of the network and the topology information are fed to the learning machine from the switches, which generates an optimal sub-graph based on the traffic volume by learning from historical data. Since the module is designed to work in a dynamic environment, low traffic load would result in a sub-graph with a smaller number of active links and switches as compared to a subgraph in the case of high traffic load. The topology manager module is responsible for retrieving information about the organization and status of the network components. It also keeps track of cost information of links and forwarding switches. If a network component fails or is out of service, the topology manager updates the global topology information.

\begin{figure}[h]
	\vspace{-.4cm}
	\centering \includegraphics [width=0.75\linewidth,trim=4 4 4 4,clip]  {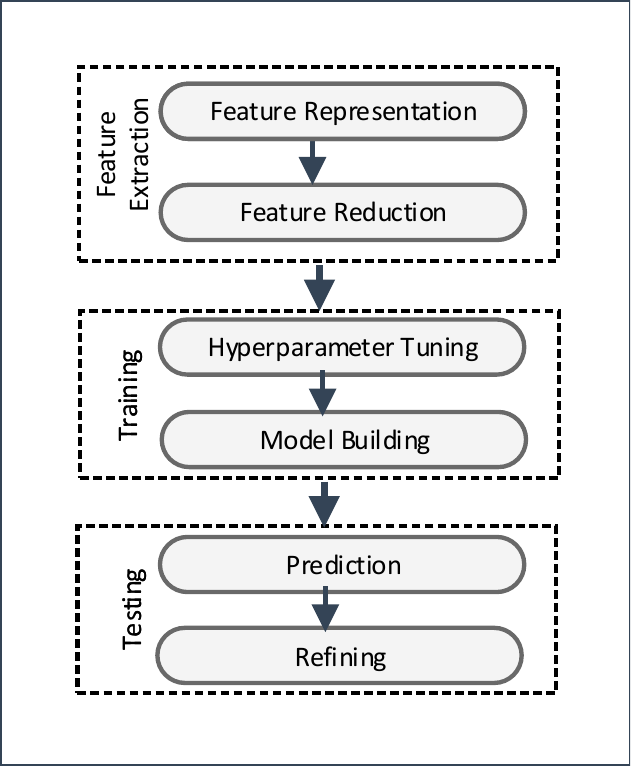}
	\caption{Inside the Learning Machine  }
	\label{fig:MLmethod} 
\end{figure}

Figure \ref{fig:MLmethod} illustrates the machine learning stages applied to learn from traffic information and statistics of the network components. The pre-processing stage extracts features from the traffic, topology, switch, and link data and represents them using a matrix to perform size reduction. The training stage sets the hyper-parameters of the training model using cross-validation, and then builds a training model. The testing stage makes a prediction on the next sub-optimal graph that is proportional to the traffic volume. The refining prediction component improves the predicted state using a heuristics algorithm.

\subsection{Feature Extraction} 

In machine learning, feature extraction is a technique used to select a subset of data more relevant to finding interesting patterns. Feature extraction involves feature representation and feature reduction. The  performance  of  machine  learning  methods depends  on the  choice of features. Complex features require memory, computational power, and longer training time. Moreover, the machine leaning algorithm over-fits the training set and generalizes poorly for unseen data.  Feature reduction is a method of reducing the dimension of the feature set. Dimension reduction is the process of reducing the number of random variables under consideration by obtaining a set of principal variables \cite{1998dimensionreduction,200nonlinear,2006foundations}. Major techniques used in machine learning are Principal Component Analysis (PCA) \cite{2006pca}, Factor Analysis (FA), Projection Pursuit (PP), and Independent Component Analysis (ICA) \cite{2002reducsurvey}. 

The network is represented as a directed graph where the nodes and the edges represent the switches and the links, respectively. A traffic flow is represented by the source node, destination node, and the flow rate. If the number of nodes in the network is N, then the size of the traffic matrix is Nx(N-1).

PCA is a linear combination of optimally-weighted observed variables. The outputs of PCA are these principal components whose numbers are less than or equal to the size of the original feature space. The principal components are orthogonal to each other. PCA is commonly used in face recognition, image classification, and unsupervised predictions. 
In this work, we use PCA to reduce the dimension of the feature set. 

\begin{algorithm}
	\caption{\textbf{PCA}: Reduce the feature data size X$^{dxn}$ to X$^{dxk}$  and produce the projection matrix W}\label{alg:pca}

\textbf{INPUT :} Feature data X$^{dxn}$ and k the number of principal components 

\textbf{OUTPUT :} Feature data X$^{dxk}$ where k is the number of principal components
\begin{algorithmic}[1] 

\State $\bar{X} \gets \sum\limits_{i=1}^n X_{i}$  \Comment{\textit{mean of X}} \label{pca:mu}

\State $\bar{X}  \gets X- \bar{X}$ \Comment{\textit{mean normalize X}} \label{pca:normalize}

\State C  $\gets \frac{1}{d}(X-\bar{X})$$^T$(X-$\bar{X}$) \Comment{\textit{C is the covariance matrix}}\label{pca:covariance}

\State {V,E} $\gets eig(C)$ \Comment{\textit{compute eigenvalues \& eigenvectors}} \label{pca:eigen}

\State V$\gets$ sort$_{desc}$(V,E)  \Comment{\textit{sort V based on E}} \label{pca:sort}

\State W $\gets$ eigenvecs$^{k}$ \Comment{Projection matrix W$^{d*k}$} \label{pca:wproject2}

\State X$^{dxk} \gets$ XW \Comment{Project X on W space} \label{pca:xprojectiononw}

\State \textbf{return} X$^{dxk},W$ \label{pca:return}
\end{algorithmic}
\end{algorithm}

Algorithm \ref{alg:pca} shows the steps used in PCA for feature size reduction. Line  \ref{pca:mu} computes the mean vector $\bar{X}$ of X. The dimension of $\bar{X}$ is equal to the feature size n. Line \ref{pca:normalize} mean normalizes the data. Mean normalization is necessary because it makes each feature component have same standard deviation which helps all principal components have equal weight. The next step in PCA is to compute the covariance matrix of the mean normalized data and compute the eigenvectors V and eigenvalues E as stipulated on lines \ref{pca:covariance} and \ref{pca:eigen} . Line \ref{pca:sort} orders the V based on eigenvalues E in descending order.  The eigenvector corresponding to the maximum eigenvalue is at the first position while the eigenvector corresponding to the minimum eigenvalue is at the end of the list. 

The next step in the PCA algorithm is to prepare the projection matrix W with the top k principal components. However, selecting the value of k is a significant step in PCA and challenging task. Small k value reduces the feature size significantly but preserves fewer information of the original data. The variance of the principal components shows the direction of the the maximum eigenvalue. The direction of the eigenvector corresponding the to maximum eigenvalue that carries most of the information in the original unreduced data. The larger the variance the more information the principal components carry. If we use the whole principal components the variance will be closer to 100\%, and if the number of principal components chosen does not carry any information about the whole matrix, the value becomes 0. The variance decreases while moving from the first (largest) component to the last one. Line \ref{pca:wproject2} computes the projection matrix W$^{nxk}$. W is used to project the original data X$^{dxn}$ or a new data sample of various size. Line \ref{pca:xprojectiononw}  reduces the dxn dimensional matrix X to dxk by projecting it over the eigenspace W. Line \ref{pca:return} returns the projected matrix X$^{dxk}$ and the projection matrix W$^{kn}$.

\subsection{Training}\label{subsec:training}
The training stage of MER-SDN has two parts: hyper-parameter tuning and model training. We use linear regression to build our training model. However, setting the number of principal components needs to be investigated carefully. We use 10 fold cross-validation to pick parameter k. In 10-fold cross-validation, the original sample is randomly partitioned into 10 equal size subsamples. Of the 10 subsamples, a single subsample is retained as the validation data for testing the model, and the remaining 9 subsamples are used as training data. The cross-validation process is then repeated 10 times (the folds), with each of the 10 subsamples used exactly once as the validation data. The 10 results from the folds can then be averaged (or otherwise combined) to produce a single estimation. The advantage of this method is that all observations are used for both training and validation, and each observation is used for validation exactly once. This avoids bias in the training. After tuning the parameters, the next step is to build a regression model by using all the training dataset.  The regression model is now ready to use for predicting new features.

\subsection{Testing}

Testing refers to applying the model trained to predict the values for unknown new data sample. First, we transform the test data set to eigenspace by a simple linear transformation. Then, we use the regression model we build (in section \ref{subsec:training}) for prediction. However, in our experiments, the accuracy of the prediction is not 100\% in constant time.

Algorithm \ref{alg:refining} increases the accuracy of the model. The rationale behind the Refine algorithm is to increase the predicted $Umin$ by  $\alpha$ until the energy saving decreases and to decrease the value of the predicted $Umax$ by  $\alpha$ until the energy saving remains constant. And the outputs are the improved $Umin$ and $Umax$. 

The inputs to the algorithm \ref{alg:refining} are predicted $Umin_{0}$, predicted $Umax_{0}$, step size $\alpha$, and threshold $\beta$. The step size parameter $\alpha$ is the value to add or subtract from $Umin$ and $Umax$ to find if the near values have better energy efficiency or not. The threshold parameter $\beta$ is the terminating condition for the algorithm that measures that measures the energy saving difference between previous and current $Umin$ and $Umax$ values.

Lines \ref{refine:eecur}, \ref{refine:eeprev}, and \ref{refine:eenext} calculate the efficiency of the energy saving algorithm MEPT for the $Umin$ parameter value of $Umin_{0}$,$Umin_{0} - \alpha$, and $Umin_{0} + \alpha$ respectively.  Lines from \ref{refine:umincheckbegin} to \ref{refine:umincheckend} if the change in $Umin$ changes the energy saving. Line \ref{refine:terminateumin} sets the terminating condition for refining the value of the predicted $Umin$ by checking if the difference between  the energy saving using the current $Umin$ and the previous $Umin$ is not greater than the threshold $\beta$.  Lines from \ref{refine:beginumax} to \ref{refine:endumax} alliteratively reduce the $Umax_{0}$ value till the energy saving is not changing according to line \ref{refine:optimalumax}. The optimal value of $Umin$ and $Umax$ are found on lines \ref{refine:optimlaumin}, and \ref{refine:optimalumax}.

\begin{algorithm}
	\caption{\textbf{Refine}: Improves the predicted parameters $Umin_{0}$ and $Umax_{0}$ values for better efficiency} \label{alg:refining}

\textbf{INPUT :} Predicted parameters $Umin_{0}$ and $Umax_{0}$, change $\alpha$, threshold $\beta$.

\textbf{OUTPUT :} Improved parameters $Umin$,$Umax$

	\begin{algorithmic}[1]  		

		\Repeat 
		\State \texttt{$EE_{curr}\gets EE(Umin_{0},Umax_{0})$} \label{refine:eecur}
		\State \texttt{$EE_{prev}\gets EE(Umin_{0}- \alpha,Umax_{0})$} \label{refine:eeprev}
		\State \texttt{$EE_{next}\gets EE(Umin_{0}+ \alpha,Umax_{0})$} \label{refine:eenext}
		\If{$EE_{prev} < EE_{next}$} \label{refine:umincheckbegin}
		\State $Umin_{0}\gets Umin_{0}+\alpha$ 
		\Else 
		\State $Umin_{0}\gets Umin_{0}-\alpha$
		\EndIf \label{refine:umincheckend}
		\State \texttt{$EE_{new}\gets EE(Umin_{0},Umax_{0})$}
 \Until{$ABS(EE_{curr}-EE_{new})\leq  \beta$} \label{refine:terminateumin} 
 \While{$EE_{curr} \geq EE(Umin_{0},Umax_{0}-\alpha)$} \label{refine:terminateumax}
 
\State\texttt{$Umax_{0} \gets Umax_{0}-\alpha$} \label{refine:beginumax}
\State \texttt{$EE_{curr}\gets EE(Umin_{0},Umax_{0})$}
 \EndWhile \label{refine:endumax}
 
 \State \texttt{$Umin \gets Umin_{0}$} \label{refine:optimlaumin}
 \State \texttt{$Umax \gets Umax_{0}$} \label{refine:optimalumax}
 \State \textbf{return} $Umin, \text{ }Umax$ \label{refine:return}
	\end{algorithmic}
\end{algorithm}

\section{Experimental Analysis}\label{sec:experiment}

The experimental platform is based on POX controller and Mininet \cite{mininet} network emulator installed on Ubuntu 16.04 64-bit. The topology is created on Mininet, and the heuristics are implemented on POX controller. Our experiments are conducted using real traces from SNDlib \cite{SNDlib10}, in particular, the Abilene, GEANT, and Nobel-Germany dynamic traffic trace of the European research network. The metrics we use in this experiment are the accuracy of the predictor, feature size reduction due to PCA, cross-fold validation to pick the optimal number of principal components, speedup of the predictor and the refine algorithm as compared to the brute force method,  energy efficiency, and average path length. Accuracy is calculated as $100*(1-\frac{|TV - PV|}{TV}) \pm \epsilon \label{eq:accuracy}$ where $TV$ is the true value of the parameter, $PV$ the predicted value of the parameter, and  $\epsilon$ is the error tolerated. In this experiment, the value of $\epsilon$ is 3\%. Speedup is calculated as $\frac{100}{N}$ where $N$ is the number of times the energy saving algorithm (MEPT) is run before the Refine algorithm gets the optimal value.

\begin{table}[ht]
\centering
\caption{Topologies and Traces}
\label{tbl:traces}
\begin{tabular}{c|ccccc}
\hline
Topology & Nodes & Edges & 
\begin{tabular}[c]{@{}l@{}}Avg \\ Deg.\end{tabular}&
\begin{tabular}[c]{@{}l@{}}Feature \\ Size\end{tabular} &
\begin{tabular}[c]{@{}l@{}}Snapshot \\Minutes\end{tabular}\\  \hline
Abilene & 12 & 15 &  2.5 &  132& 5 \\
GEANT & 22 & 36  & 4.35 & 462 & 15 \\
Nobel-Germany & 17 & 26 & 3.06& 272  & 5 \\ \hline \hline
\end{tabular}
\end{table}

\begin{algorithm}
	\caption{\textbf{PCA}: Reduce the traffic matrix X$^{nxd}$ to X$^{nxk}$  and produce the projection matrix W$^{dxk}$}\label{alg:pca}

\textbf{Input:} Traffic matrix X$^{nxd}$ and k the number of principal components 

\textbf{Output:} Feature data X$^{nxk}$ and projection matrix W$^{dxk}$ where k is the number of principal components and  d is  

\begin{algorithmic}[1] 
\State $\bar{X} \gets \sum\limits_{i=1}^d X_{i}$  \Comment{mean of X} \label{pca:mu}
\State $T \gets $X - $\bar{X}$ \Comment{mean normalize T} \label{pca:normalize}
\State $C \gets \frac{1}{n}(X-\bar{X})^{T}(X-\bar{X})$ \Comment{C is the covariance matrix}\label{pca:covariance}
\State $V,E \gets eig(C)$\label{pca:eigen} \Comment{ computer eigen value and vector}
\State $V \gets sort_{desc}(V,E)$ \Comment{sort V based on E} \label{pca:sort}
\State $W \gets eigenvecs^{k}$ \Comment{Projection matrix $W^{dxk}$} \label{pca:wproject}
\State $X^{nxk} \gets XW$ \Comment{Project X on W space} \label{pca:xprojectiononwx}
\end{algorithmic}
\end{algorithm}

Table \ref{tbl:traces} presents the topologies, traffic characteristic and the size of the features used in this experiment. The features are extracted from a snapshot of the network aggregated in 5 to 15 minutes. The features are represented as a matrix where each row is an N(N-1) vector representing the rates between a source and destination pairs. For the Abilene topology with 12 nodes, accordingly, the dimension of the feature is 12*11=132. Accordingly, the feature sizes for GEANT and Nobel-Germany are 462 and 272 respectively. We train the models for traffic size of 10\% to 90\%. 

\begin{algorithm}
\caption{MaxRESDN ($\mathbb{G}$, $\mathbb{F}$,$\mathbb{U}$,$Umin$,$Umax$)}\label{alg:MaxRESDN }
	\begin{algorithmic}[1]  		
		\State \textbf{Input: } Graph $\mathbb{G}$, set of traffic flow  $\mathbb{F}$, utility of links $\mathbb{U}$, minimum utility $Umin$, and maximum utility $Umax$
	\State \textbf{Output: } Modified utility of links $\mathbb{U}$ and graph $\mathbb{G}$
		\ForAll{\texttt{$f=(sr,ds,\lambda_{f}) \in F$}} \label{RESDN :flow}
		\State \texttt{$path_{f} \gets$  PathMaxRESDN (sr,ds,$\lambda_{f})$} \label{RESDN :pathmaxi}
		\ForAll{\texttt{$e_{ij} \in path_{f}$}} 		
		\State \texttt{$U_{ij} \gets  U_{ij}+\dfrac{\lambda_{f}}{W_{ij}}$}  \label{RESDN :Uupdate}
		\EndFor  
		\EndFor
		\ForAll{\texttt{$e_{ij} \in \mathbb{E}$}} \label{RESDN :updategraph}		
		\If{$U_{ij} == 0$}
    		    \State $L_{ij} \gets  0$ 
    		   \EndIf
		\EndFor \label{RESDN :updatgraphend}
	\end{algorithmic}
\end{algorithm}

\begin{algorithm}
	\caption{\textbf{PCA}: Reduce the traffic matrix X$^{nxd}$ to X$^{nxk}$  and produce the projection matrix W$^{dxk}$}\label{alg:pca}

\textbf{Input :} Traffic matrix X$^{nxd}$ and k the number of principal components 

\textbf{Output :} Feature data X$^{nxk}$ and projection matrix W$^{dxk}$ where k is the number of principal components and  d is  
\begin{algorithmic}[1] 
\State $\bar{X} \gets \sum\limits_{i=1}^d X_{i}$  \Comment{\textit{mean of X}} \label{pca:mu}
\State $T \gets $X - $\bar{X}$ \Comment{mean normalize T} \label{pca:normalize}
\State $C \gets \frac{1}{n}(X-\bar{X})^{T}(X-\bar{X})$ \Comment{\textit{C is the covariance matrix}}\label{pca:covariance}
\State $V,E \gets eig(C)$\label{pca:eigen} \Comment{ computer eigen value and vector}
\State $V \gets sort_{desc}(V,E)$ \Comment{sort V based on E} \label{pca:sort}
\State $W \gets eigenvecs^{k}$ \Comment{Projection matrix $W^{dxk}$} \label{pca:wproject}
\State $X^{nxk} \gets XW$ \Comment{Project X on W space} \label{pca:xprojectiononw}
\end{algorithmic}
\end{algorithm}

\begin{figure}[ht]
	\centering
	\begin{minipage}[b]{.71\textwidth}
		\scalebox{0.55}{
%
%
\definecolor{mycolor1}{rgb}{0.00000,0.44700,0.74100}%
\definecolor{mycolor2}{rgb}{0.85000,0.32500,0.09800}%
\definecolor{mycolor3}{rgb}{0.92900,0.69400,0.12500}%
\definecolor{mycolor4}{rgb}{0.49400,0.18400,0.55600}%
\definecolor{mycolor250}{rgb}{0.6400,0.1400,0.35600}
\definecolor{mycolorblack}{rgb}{0,0,0}
\definecolor{mycolor5}{rgb}{0.46600,0.67400,0.18800}%
\definecolor{mycolor6}{rgb}{0.30100,0.74500,0.93300}%
\begin{tikzpicture}

\begin{axis}[%
width=4.521in,
height=3.566in,
at={(1.322in,0.742in)},
scale only axis,
xmin=10,
xmax=100,
ymin=0,
ymax=100,
ylabel style={font=\color{white!15!black}},
ylabel={Percentage},
xlabel style={font=\large},
xlabel={Principal Components (\%)},
ylabel style={font=\large},
axis background/.style={fill=white},
axis x line*=bottom,
axis y line*=left,
xmajorgrids,
ymajorgrids,
legend style={at={(0.582,0.502)}, anchor=south west, legend columns=1, legend cell align=left, align=left, draw=white!15!black}
]

\addplot [smooth, color=mycolorblack, line width=1.5pt, mark size=3pt, mark=triangle, mark options={solid, mycolorblack}]
  table[row sep=crcr]{%
10.82251082	89.177\\
16.23376623	80.177\\
21.64502165	78.35\\
27.05627706	74.34\\
32.46753247	67.53\\
37.87878788	63.45\\
43.29004329	56.71\\
48.7012987	53.99\\
54.11255411	45.88\\
64.93506494	35.01\\
75.75757576	24.24\\
86.58008658	13.42\\
97.4025974	2.6\\
};
\addlegendentry{Size Reduction}

\addplot [smooth, color=mycolor1, line width=1.5pt, mark size=3pt, mark=otimes, mark options={solid, mycolor1}]
  table[row sep=crcr]{%
10.82251082	37.177\\
16.23376623	42.177\\
21.64502165	54.35\\
27.05627706	77.34\\
32.46753247	80.53\\
37.87878788	81.45\\
43.29004329	83.71\\
48.7012987	84.01\\
54.11255411	83.01\\
64.93506494	84.01\\
75.75757576	83.01\\
86.58008658	83.01\\
97.4025974	83.01\\
};
\addlegendentry{Accuracy}

\addplot [smooth, color=mycolor2, line width=1.5pt, mark size=3pt, mark=square, mark options={solid, mycolor2}]
  table[row sep=crcr]{%
10.82251082	54\\
21.64502165	76\\
32.46753247	88\\
43.29004329	92\\
54.11255411	96\\
64.93506494	95\\
75.75757576	96.5\\
86.58008658	96\\
97.4025974	98.99\\
};
\addlegendentry{Variance}
\addplot [dotted, color=black, line width=0pt, mark size=5pt, mark=x, mark options={solid, black}]
  table[row sep=crcr]{%
30.01	0\\
30.01	10\\
30.01	20\\
30.01	30\\
30.01	40\\
30.01	50\\
30.01	60\\
30.01	70\\
30.01	80\\
30.01	90\\
30.01	100\\
};

\end{axis}
\end{tikzpicture}%
} \subcaption{Abilene}\label{fig:crossabiline}
	\end{minipage}\qquad
	\hspace{-1.8cm}
	\begin{minipage}[b]{.71\textwidth}
		\scalebox{0.55}{
%
%
\definecolor{mycolor1}{rgb}{0.00000,0.44700,0.74100}%
\definecolor{mycolor2}{rgb}{0.85000,0.32500,0.09800}%
\definecolor{mycolor3}{rgb}{0.92900,0.69400,0.12500}%
\definecolor{mycolor4}{rgb}{0.49400,0.18400,0.55600}%
\definecolor{mycolor250}{rgb}{0.6400,0.1400,0.35600}
\definecolor{mycolorblack}{rgb}{0,0,0}
\definecolor{mycolor5}{rgb}{0.46600,0.67400,0.18800}%
\definecolor{mycolor6}{rgb}{0.30100,0.74500,0.93300}%
\begin{tikzpicture}

\begin{axis}[%
width=4.521in,
height=3.566in,
at={(1.322in,0.742in)},
scale only axis,
xmin=10,
xmax=100,
ymin=0,
ymax=100,
ylabel style={font=\color{white!15!black}},
ylabel={Percentage},
xlabel style={font=\large},
xlabel={Principal Components (\%)},
ylabel style={font=\large},
axis background/.style={fill=white},
axis x line*=bottom,
axis y line*=left,
xmajorgrids,
ymajorgrids,
legend style={at={(0.582,0.502)}, anchor=south west, legend columns=1, legend cell align=left, align=left, draw=white!15!black}
]
\addplot [smooth, color=mycolorblack, line width=1.5pt, mark size=3pt, mark=triangle, mark options={solid, mycolorblack}]
  table[row sep=crcr]{%
10.82251082	89.177\\
16.23376623	80.177\\
21.64502165	78.35\\
27.05627706	74.34\\
32.46753247	67.53\\
37.87878788	63.45\\
43.29004329	56.71\\
48.7012987	53.99\\
54.11255411	45.88\\
64.93506494	35.01\\
75.75757576	24.24\\
86.58008658	13.42\\
97.4025974	2.6\\
};
\addlegendentry{Size Reduction}

\addplot [smooth, color=mycolor1, line width=1.5pt, mark size=3pt, mark=otimes, mark options={solid, mycolor1}]
  table[row sep=crcr]{%
10.82251082	37.177\\
16.23376623	43.177\\
21.64502165	55.35\\
27.05627706	74.34\\
32.46753247	80.53\\
37.87878788	82.45\\
43.29004329	86.71\\
48.7012987	88.01\\
54.11255411	88.01\\
64.93506494	88.01\\
75.75757576	88.01\\
86.58008658	88.01\\
97.4025974	88.01\\
};
\addlegendentry{Accuracy}

\addplot [smooth, color=mycolor2, line width=1.5pt, mark size=3pt, mark=square, mark options={solid, mycolor2}]
  table[row sep=crcr]{%
10.82251082	56\\
21.64502165	78\\
32.46753247	88\\
43.29004329	92\\
54.11255411	94\\
64.93506494	97\\
75.75757576	97.5\\
86.58008658	98\\
97.4025974	98.99\\
};
\addlegendentry{Variance}
\addplot [dotted, color=black, line width=0pt, mark size=5pt, mark=x, mark options={solid, black}]
  table[row sep=crcr]{%
32.01	0\\
32.01	10\\
32.01	20\\
32.01	30\\
32.01	40\\
32.01	50\\
32.01	60\\
32.01	70\\
32.01	80\\
32.01	90\\
32.01	100\\
};

\end{axis}
\end{tikzpicture}%
} \subcaption{GEANT}\label{fig:crossgeant}
	\end{minipage}\qquad
	\hspace{-1.8cm}
	\begin{minipage}[b]{.71\textwidth}
		\scalebox{0.55}{
%
%
\definecolor{mycolor1}{rgb}{0.00000,0.44700,0.74100}%
\definecolor{mycolor2}{rgb}{0.85000,0.32500,0.09800}%
\definecolor{mycolor3}{rgb}{0.92900,0.69400,0.12500}%
\definecolor{mycolor4}{rgb}{0.49400,0.18400,0.55600}%
\definecolor{mycolor250}{rgb}{0.6400,0.1400,0.35600}
\definecolor{mycolorblack}{rgb}{0,0,0}
\definecolor{mycolor5}{rgb}{0.46600,0.67400,0.18800}%
\definecolor{mycolor6}{rgb}{0.30100,0.74500,0.93300}%
\begin{tikzpicture}

\begin{axis}[%
width=4.521in,
height=3.566in,
at={(1.322in,0.742in)},
scale only axis,
xmin=10,
xmax=100,
ymin=0,
ymax=100,
ylabel style={font=\color{white!15!black}},
ylabel={Percentage},
xlabel style={font=\large},
xlabel={Principal Components (\%)},
ylabel style={font=\large},
axis background/.style={fill=white},
axis x line*=bottom,
axis y line*=left,
xmajorgrids,
ymajorgrids,
legend style={at={(0.582,0.502)}, anchor=south west, legend columns=1, legend cell align=left, align=left, draw=white!15!black}
]
\addplot [smooth, color=mycolorblack, line width=1.5pt, mark size=3pt, mark=triangle, mark options={solid, mycolorblack}]
  table[row sep=crcr]{%
11.02941176	88.9705882352941\\
22.05882353	77.9411764705882\\
33.08823529	66.9117647058823\\
44.11764706	55.8823529411765\\
55.14705882	44.8529411764706\\
66.17647059	33.8235294117647\\
77.20588235	22.7941176470588\\
88.23529412	11.7647058823529\\
99.26470588	0.735294117647058\\
};
\addlegendentry{Size Reduction}

\addplot [smooth, color=mycolor1, line width=1.5pt, mark size=3pt, mark=otimes, mark options={solid, mycolor1}]
  table[row sep=crcr]{%
11.02941176	43.177\\
22.05882353	47.177\\
33.08823529	59.35\\
44.11764706	80.34\\
55.14705882	82.53\\
66.17647059	84.45\\
77.20588235	88.71\\
82.23529412	91.01\\
99.26470588	91.01\\
};
\addlegendentry{Accuracy}

\addplot [smooth, color=mycolor2, line width=1.5pt, mark size=3pt, mark=square, mark options={solid, mycolor2}]
  table[row sep=crcr]{%
11.02941176	53\\
22.05882353	71\\
33.08823529	83\\
44.11764706	87\\
55.14705882	91\\
66.17647059	91\\
77.20588235	93.5\\
88.23529412	96\\
99.26470588	98.9\\
};
\addlegendentry{Variance}
\addplot [dotted, color=black, line width=0pt, mark size=5pt, mark=x, mark options={solid, black}]
  table[row sep=crcr]{%
35.01	0\\
35.01	10\\
35.01	20\\
35.01	30\\
35.01	40\\
35.01	50\\
35.01	60\\
35.01	70\\
35.01	80\\
35.01	90\\
35.01	100\\
};

\end{axis}
\end{tikzpicture}%
}
		\subcaption{Nobel-Germany}\label{fig:crossnobelgermany}
	\end{minipage}\qquad
	\caption{Cross validation results of the percentage of PCs versus feature  size reduction, accuracy, and variance for Abilene, GEANT, and  Nobel-Germany topology traces. It also shows the optimal percentage of PCs selected for each topology and trace.}
	\label{fig:crossvalidation}	
\end{figure}
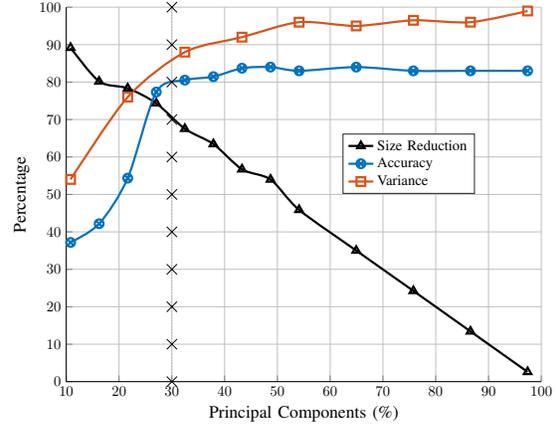
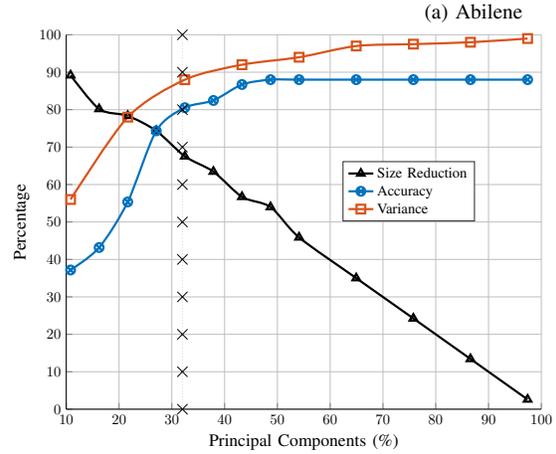
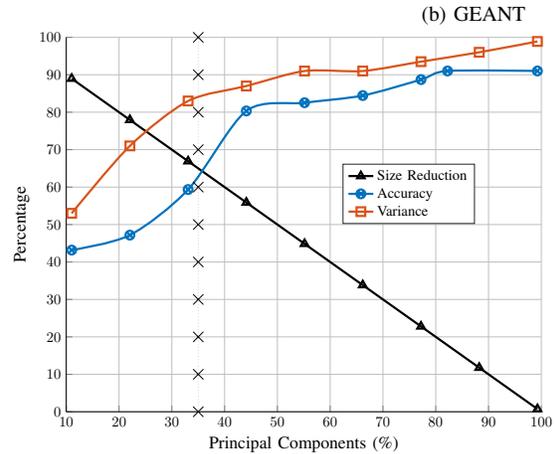

Figure \ref{fig:crossvalidation} show size reduction, cross-validation accuracy and variance of the model for the Abilene, GEANT, and Nobel-Germany topology and trace. Size reduction is inversely proportional to accuracy and variance. The percentage of principal components we picked from the 10-fold cross validation are  30\%, 32\% and 35\% which correspond to accuracy values  78\%, 79\%, 80\% and feature size reduction of 70\%, 68\%, 65\% for the Abilene, GEANT and Nobel-Germany topologies and traces. The value of k we have chosen for the PCA is 40, 148, and 96 for the three topologies and traces. The number of principal components we picked is up to 5\% larger than where the accuracy and the size reduction plots intersect. The choice is carefully made so that the model would not over-fit the data at the same time contain at least 80\% of the information in the original data.

\begin{figure}[ht]
	\centering
	\begin{minipage}[b]{.71\textwidth}
		\scalebox{0.55}{
%
%
\definecolor{mycolor1}{rgb}{0,0.7490,1}%
\definecolor{mycolor2}{rgb}{0.2745,0.5098,0.70588}%
\definecolor{mycolor3}{rgb}{0,0,0.5019}%
\definecolor{mycolor4}{rgb}{0,0,0.8039}%
\definecolor{mycolor5}{rgb}{0.75294,0.75294,0.75294}%
\definecolor{mycolor6}{rgb}{0.50196,0.50196,0.50196}%
\definecolor{mycolor7}{rgb}{0.412,0.412,0.412}
\definecolor{mycolor8}{rgb}{0,0,0}
\begin{tikzpicture}

\begin{axis}[%
width=4.521in,
height=3.566in,
at={(1.322in,0.742in)},
scale only axis,
xmin=10,
xmax=100,
ymin=40,
ymax=100,
ytick={10,20,30,40,50,60,70,80,90,100},
xtick={10,20,30,40,50,60,70,80,90,100},
ylabel style={font=\color{white!15!black}},
ylabel={Accuracy (\%)},
xlabel style={font=\large},
xlabel={Traffic Volume (\%)},
ylabel style={font=\large},
axis background/.style={fill=white},
axis x line*=bottom,
axis y line*=left,
xmajorgrids,
ymajorgrids,
legend style={at={(0.10,0.789)}, anchor=south west, legend columns=2, legend cell align=left, align=left,font=\Large draw=white!15!black}
]

\addplot [color=green, line width=1.5pt, mark size=3.0pt, mark=diamond, mark options={solid, green}]
 plot [error bars/.cd, y dir = both, y explicit]
 table[row sep=crcr, y error plus index=2, y error minus index=3]{%
10	75	2	2\\
20	71	1	1\\
30	69	2	2\\
40	72	3	3\\
50	73	3	3\\
60	71	3	3\\
70	70	1	1\\
80	71	1	1\\
90	68	2	2\\
};
\addlegendentry{Umin}
\addplot [color=mycolor8, line width=1.5pt, mark size=3.0pt, mark=triangle, mark options={solid, mycolor8}]
 plot [error bars/.cd, y dir = both, y explicit]
 table[row sep=crcr, y error plus index=2, y error minus index=3]{%
10	78	2	2\\
20	73	3	3\\
30	73	2	2\\
40	76	1	1\\
50	74	1	1\\
60	75	2	2\\
70	71	3	3\\
80	72	3	3\\
90	69	3	3\\
};
\addlegendentry{Umin/Umax}
\addplot [color=orange, line width=1.5pt, mark size=3.0pt, mark=otimes, mark options={solid, orange}]
 plot [error bars/.cd, y dir = both, y explicit]
 table[row sep=crcr, y error plus index=2, y error minus index=3]{%
10	77	2	2\\
20	78	3	3\\
30	76	1	1\\
40	75	1	1\\
50	75	3	3\\
60	75	2	2\\
70	80	2	2\\
80	76	1	1\\
90	76	2	2\\
};
\addlegendentry{Umax}
\addplot [color=mycolor4, line width=1.5pt, mark size=3.0pt, mark=diamond, mark options={solid, mycolor4}]
 plot [error bars/.cd, y dir = both, y explicit]
 table[row sep=crcr, y error plus index=2, y error minus index=3]{%
10	81	1	1\\
20	81	1	1\\
30	78	3	3\\
40	79	1	1\\
50	76	1	1\\
60	76	1	1\\
70	83	1	1\\
80	77	1	1\\
90	78	3	3\\
};
\addlegendentry{Umax/Umin}

\end{axis}
\end{tikzpicture}%
} \subcaption{Abilene}\label{fig:accuracyabiline}
	\end{minipage}\qquad
	\hspace{-1.8cm}
	\begin{minipage}[b]{.71\textwidth}
		\scalebox{0.55}{
%
%
\definecolor{mycolor1}{rgb}{0,0.7490,1}%
\definecolor{mycolor2}{rgb}{0.2745,0.5098,0.70588}%
\definecolor{mycolor3}{rgb}{0,0,0.5019}%
\definecolor{mycolor4}{rgb}{0,0,0.8039}%
\definecolor{mycolor5}{rgb}{0.75294,0.75294,0.75294}%
\definecolor{mycolor6}{rgb}{0.50196,0.50196,0.50196}%
\definecolor{mycolor7}{rgb}{0.412,0.412,0.412}
\definecolor{mycolor8}{rgb}{0,0,0}
\begin{tikzpicture}

\begin{axis}[%
width=4.521in,
height=3.566in,
at={(1.322in,0.742in)},
scale only axis,
xmin=10,
xmax=100,
ymin=40,
ymax=100,
ytick={10,20,30,40,50,60,70,80,90,100},
xtick={10,20,30,40,50,60,70,80,90,100},
ylabel style={font=\color{white!15!black}},
ylabel={Accuracy (\%)},
xlabel style={font=\large},
xlabel={Traffic Volume (\%)},
ylabel style={font=\large},
axis background/.style={fill=white},
axis x line*=bottom,
axis y line*=left,
xmajorgrids,
ymajorgrids,
legend style={at={(0.30,0.135)}, anchor=south west, legend columns=2, legend cell align=left, align=left,font=\Large draw=white!15!black}
]
\addplot [color=green, line width=1.5pt, mark size=3.0pt, mark=diamond, mark options={solid, green}]
 plot [error bars/.cd, y dir = both, y explicit]
 table[row sep=crcr, y error plus index=2, y error minus index=3]{%
10	74	3	3\\
20	72	2	2\\
30	74	1	1\\
40	76	3	3\\
50	71	3	3\\
60	72	2	2\\
70	74	2	2\\
80	76	3	3\\
90	77	1	1\\
};
\addlegendentry{Umin}
\addplot [color=orange, line width=1.5pt, mark size=3.0pt, mark=otimes, mark options={solid, orange}]
 plot [error bars/.cd, y dir = both, y explicit]
 table[row sep=crcr, y error plus index=2, y error minus index=3]{%
10	70	2	2\\
20	72	1	1\\
30	72	2	2\\
40	75	1	1\\
50	76	2	2\\
60	72	2	2\\
70	79	3	3\\
80	78	1	1\\
90	79	1	1\\
};
\addlegendentry{Umax}
\addplot [color=mycolor4, line width=1.5pt, mark size=3.0pt, mark=diamond, mark options={solid, mycolor4}]
 plot [error bars/.cd, y dir = both, y explicit]
 table[row sep=crcr, y error plus index=2, y error minus index=3]{%
10	92	2	2\\
20	94	3	3\\
30	89	3	3\\
40	91	2	2\\
50	96	2	2\\
60	96	3	3\\
70	92	1	1\\
80	92	3	3\\
90	95	1	1\\
};
\addlegendentry{Umax/Umin}
\addplot [color=mycolor8, line width=1.5pt, mark size=3.0pt, mark=triangle, mark options={solid, mycolor8}]
 plot [error bars/.cd, y dir = both, y explicit]
 table[row sep=crcr, y error plus index=2, y error minus index=3]{%
10	75	2	2\\
20	81	2	2\\
30	74	1	1\\
40	80	3	3\\
50	81	2	2\\
60	77	2	2\\
70	76	2	2\\
80	82	1	1\\
90	76	3	3\\
};
\addlegendentry{Umin/Umax}

\end{axis}
\end{tikzpicture}%
} \subcaption{GEANT}\label{fig:accuracygeant}
	\end{minipage}\qquad
	\hspace{-1.8cm}
	\begin{minipage}[b]{.71\textwidth}
		\scalebox{0.55}{
%
%
\definecolor{mycolor1}{rgb}{0,0.7490,1}%
\definecolor{mycolor2}{rgb}{0.2745,0.5098,0.70588}%
\definecolor{mycolor3}{rgb}{0,0,0.5019}%
\definecolor{mycolor4}{rgb}{0,0,0.8039}%
\definecolor{mycolor5}{rgb}{0.75294,0.75294,0.75294}%
\definecolor{mycolor6}{rgb}{0.50196,0.50196,0.50196}%
\definecolor{mycolor7}{rgb}{0.412,0.412,0.412}
\definecolor{mycolor8}{rgb}{0,0,0}
\begin{tikzpicture}

\begin{axis}[%
width=4.521in,
height=3.566in,
at={(1.322in,0.742in)},
scale only axis,
xmin=10,
xmax=100,
ymin=40,
ymax=100,
ytick={10,20,30,40,50,60,70,80,90,100},
xtick={10,20,30,40,50,60,70,80,90,100},
ylabel style={font=\color{white!15!black}},
ylabel={Accuracy (\%)},
xlabel style={font=\large},
xlabel={Traffic Volume (\%)},
ylabel style={font=\large},
axis background/.style={fill=white},
axis x line*=bottom,
axis y line*=left,
xmajorgrids,
ymajorgrids,
legend style={at={(0.10,0.0389)}, anchor=south west, legend columns=2, legend cell align=left, align=left,font=\Large draw=white!15!black}
]
\addplot [color=green, line width=1.5pt, mark size=3.0pt, mark=diamond, mark options={solid, green}]
 plot [error bars/.cd, y dir = both, y explicit]
 table[row sep=crcr, y error plus index=2, y error minus index=3]{%
10	68	3	3\\
20	74	1	1\\
30	72	3	3\\
40	77	3	3\\
50	75	3	3\\
60	74	3	3\\
70	69	1	1\\
80	68	1	1\\
90	72	2	2\\
};
\addlegendentry{Umin}
\addplot [color=mycolor8, line width=1.5pt, mark size=3.0pt, mark=triangle, mark options={solid, mycolor8}]
 plot [error bars/.cd, y dir = both, y explicit]
 table[row sep=crcr, y error plus index=2, y error minus index=3]{%
10	73	2	2\\
20	77	1	1\\
30	77	2	2\\
40	79	2	2\\
50	79	3	3\\
60	79	1	1\\
70	70	1	1\\
80	73	2	2\\
90	74	3	3\\
};
\addlegendentry{Umin/Umax}
\addplot [color=orange, line width=1.5pt, mark size=3.0pt, mark=otimes, mark options={solid, orange}]
 plot [error bars/.cd, y dir = both, y explicit]
 table[row sep=crcr, y error plus index=2, y error minus index=3]{%
10	74	1	1\\
20	72	3	3\\
30	71	3	3\\
40	75	2	2\\
50	75	1	1\\
60	77	1	1\\
70	74	1	1\\
80	71	2	2\\
90	80	2	2\\
};
\addlegendentry{Umax}
\addplot [color=mycolor4, line width=1.5pt, mark size=3.0pt, mark=diamond, mark options={solid, mycolor4}]
 plot [error bars/.cd, y dir = both, y explicit]
 table[row sep=crcr, y error plus index=2, y error minus index=3]{%
10	75	3	3\\
20	77	1	1\\
30	74	3	3\\
40	78	1	1\\
50	77	2	2\\
60	80	2	2\\
70	76	3	3\\
80	74	2	2\\
90	82	2	2\\
};
\addlegendentry{Umax/Umin}

\end{axis}
\end{tikzpicture}%
}
		\subcaption{Nobel-Germany}\label{fig:accuracynobelgermany}
	\end{minipage}\qquad
	\caption{Accuracy for predicting $Umin$, $Umax$, $Umin$/$Umax$ ($Umin$ given $Umax$ is known, and $Umin$/$Umax$ ($Umax$ given $Umin$ is known a) Abilene  b) GEANT  and c) Nobel-Germany topology traces}
	\label{fig:accuracy}	
\end{figure}
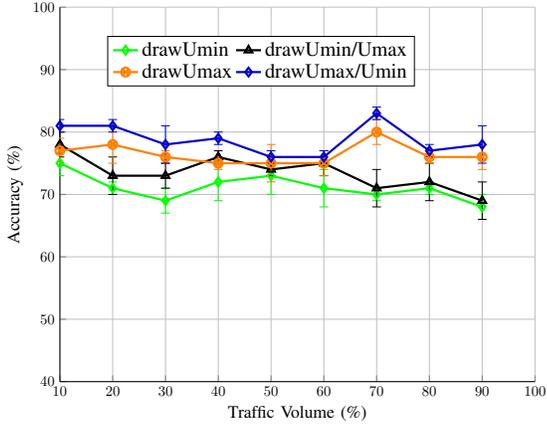
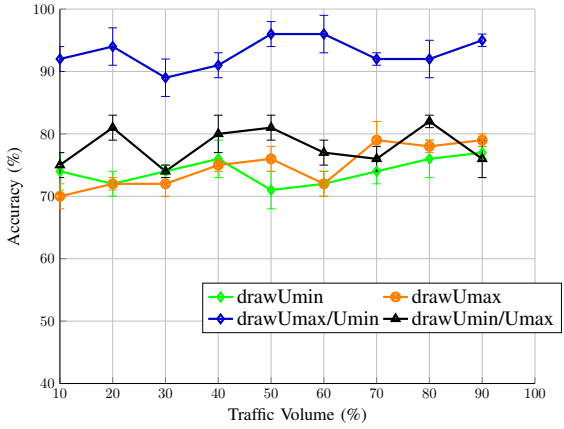
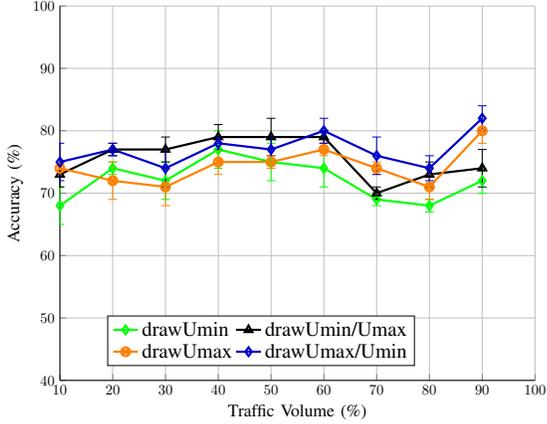

Figure \ref{fig:accuracy} shows the accuracy of predicting $Umin$, $Umax$, $Umin/Umax$ ($Umax$ given $Umin$ is known), $Umax/Umin$ ($Umin$ given $Umax$ is known) versus for the GEANT data set. The accuracy of predicting $Umin$ ranges between 68\% to 75\%. For the Abilene trace, $Umax$ prediction accuracy is 3 to 5\% better than $Umin$'s prediction. An interesting observation from this experiment is that the accuracy of $Umin$ and $Umax$ increase if $Umax$ and $Umin$ are known apriori. In case of the GEANT topology trace, a prior knowledge of $Umin$ increases the prediction accuracy of $Umax$ by at least 15\%. The accuracy of the predictor is independent of the traffic volume.

\begin{table}[ht]
\centering
\caption{Refine algorithm speedup for convergence of the  $Umin$ and $Umax$ parameters of MEPT heuristics algorithm as compared to the brute force method}
\label{tbl:speedup}
\begin{tabular}{l|ll|ll|ll}
\hline \hline
Traffic&Abiline&&GEANT&&Nobel-Germany\\
& $Umin$& $Umax$& $Umin$ & $Umax$ & $Umin$& $Umax$\\ \hline
10& 17.86 & 21.43 & 18.37& 15.43& 14.29& 18.37\\
20& 16.07 & 22.69 & 16.77& 16.77& 18.37& 16.77\\
30& 14.84 & 20.3  & 18.37& 16.77& 16.77& 16.07\\
40& 16.77 & 19.29 & 20.3 & 19.29& 21.43& 19.29\\
50& 17.53 & 19.29 & 16.07& 20.3 & 19.29& 19.29\\
60& 16.07 & 19.29 & 16.77& 16.77& 18.37& 21.43\\
70& 15.43 & 25.71 & 18.37& 24.11& 14.84& 18.37\\
80& 16.07 & 20.3  & 20.3 & 22.69& 14.29& 16.07\\
90& 14.29 & 20.3  & 21.43& 24.11& 16.77& 25.71 \\ \hline \hline  
\end{tabular}
\end{table}

Table \ref{tbl:speedup} shows fast the Refine algorithm converges the optimal values of $Umin$ and $Umax$ parameters relative to the brute force method. The brute force method checks all values from 0\% to 100 \% and selects the optimal $Umin$ and $Umax$ for highest energy saving. Since the accuracy of the prediction is not 100\%, the Refine heuristic improves the predicted values to reach the optimal value with few numbers of steps. The speedup for traffic ranging from 10\% to 90\% traffic volume is 14.85X to 25.71X the brute force method. Like the accuracy of the predictor, the speedup of the Refine heuristics is independent of the traffic volume.

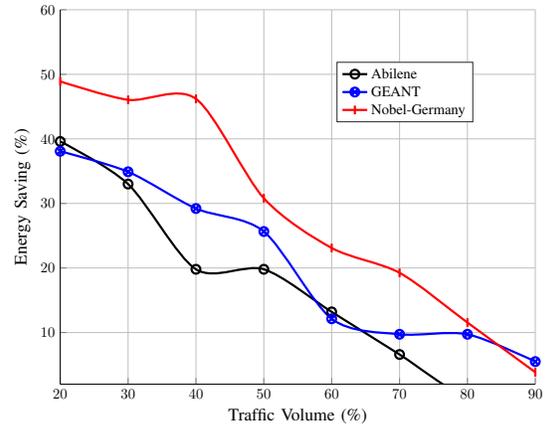
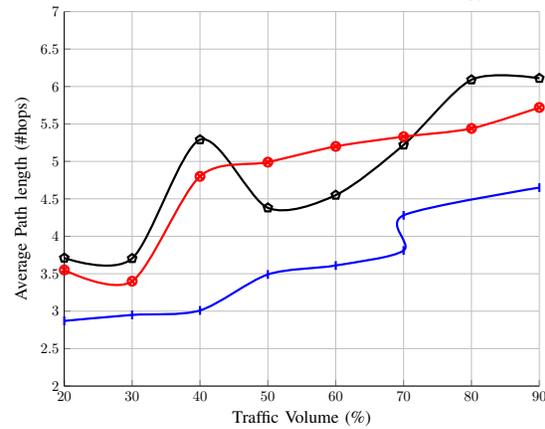
\begin{figure}[h]
	\begin{minipage}[b]{.71\textwidth}
		\scalebox{0.55}{
%
%
\definecolor{mycolor1}{rgb}{0,0.7490,1}%
\definecolor{mycolor2}{rgb}{0.2745,0.5098,0.70588}%
\definecolor{mycolor3}{rgb}{0,0,0.5019}%
\definecolor{mycolor4}{rgb}{0,0,0.8039}%
\definecolor{mycolor5}{rgb}{0.75294,0.75294,0.75294}%
\definecolor{mycolor6}{rgb}{0.50196,0.50196,0.50196}%
\definecolor{mycolor7}{rgb}{0.412,0.412,0.412}
\definecolor{mycolor8}{rgb}{0,0,0}
\begin{tikzpicture}

\begin{axis}[%
width=4.521in,
height=3.566in,
at={(1.322in,0.742in)},
scale only axis,
xmin=20,
xmax=90,
ymin=2,
ymax=60,
ylabel style={font=\color{white!15!black}},
ylabel={Energy Saving (\%)},
xlabel style={font=\large},
xlabel={Traffic Volume (\%)},
ylabel style={font=\large},
axis background/.style={fill=white},
axis x line*=bottom,
axis y line*=left,
xmajorgrids,
ymajorgrids,
legend style={at={(0.582,0.7)}, anchor=south west, legend columns=1, legend cell align=left, align=left, draw=white!15!black}
]
\addplot [smooth, color=black, line width=1.5pt, mark size=3.0pt, mark=o, mark options={solid, black}]
  table[row sep=crcr]{%
20	39.6\\
30	33\\
40	19.8\\
50	19.8\\
60	13.2\\
70	6.6\\
80	0\\
90	0\\
};
\addlegendentry{Abilene}

\addplot [smooth, color=blue, line width=1.5pt, mark size=3.0pt, mark=otimes, mark options={solid, blue}]
  table[row sep=crcr]{%
20	38.08\\
30	34.90\\
40	29.21\\
50	25.65\\
60	12.13\\
70	9.73\\
80	9.73\\
90	5.5\\
};
\addlegendentry{GEANT}
\addplot [smooth, color=red, line width=1.5pt, mark size=3.0pt, mark=|, mark options={solid, red}]
  table[row sep=crcr]{%
20	48.9\\
30	46.05\\
40	46.2\\
50	30.8\\
60	23.1\\
70	19.25\\
80	11.55\\
90	3.85\\
};
\addlegendentry{Nobel-Germany}

\end{axis}
\end{tikzpicture}%
} \subcaption{Energy Saving}\label{fig:ES}
	\end{minipage}\qquad
	\hspace{-1.706cm}
	\begin{minipage}[b]{.71\textwidth}
		\scalebox{0.55}{
%
%
\definecolor{mycolor1}{rgb}{0,0.7490,1}%
\definecolor{mycolor2}{rgb}{0.2745,0.5098,0.70588}%
\definecolor{mycolor3}{rgb}{0,0,0.5019}%
\definecolor{mycolor4}{rgb}{0,0,0.8039}%
\definecolor{mycolor5}{rgb}{0.75294,0.75294,0.75294}%
\definecolor{mycolor6}{rgb}{0.50196,0.50196,0.50196}%
\definecolor{mycolor7}{rgb}{0.412,0.412,0.412}
\definecolor{mycolor8}{rgb}{0,0,0}
\begin{tikzpicture}

\begin{axis}[%
width=4.521in,
height=3.566in,
at={(1.322in,0.742in)},
scale only axis,
xmin=20,
xmax=90,
ymin=2,
ymax=7,
ylabel style={font=\color{white!15!black}},
ylabel={Average Path length (\#hops)},
xlabel style={font=\large},
xlabel={Traffic Volume (\%)},
ylabel style={font=\large},
axis background/.style={fill=white},
axis x line*=bottom,
axis y line*=left,
xmajorgrids,
ymajorgrids,
legend style={at={(0.03,0.45)}, anchor=south west, legend columns=1, legend cell align=left, align=left, draw=white!15!black}
]
\addplot [smooth, color=black, line width=1.5pt, mark size=3.0pt, mark=pentagon, mark options={solid, black}]
  table[row sep=crcr]{%
20	3.71\\
30	3.704\\
40	5.29\\
50	4.38\\
60	4.55\\
70	5.22\\
80	6.09\\
90	6.11\\
};

\addplot [smooth, color=blue, line width=1.5pt, mark size=3.0pt, mark=|, mark options={solid, blue}]
  table[row sep=crcr]{%
20	2.87\\
30	2.95\\
40	3.01\\
50	3.49\\
60	3.61\\
70	3.81\\
70	4.28\\
90	4.65\\
};

\addplot [smooth, color=red, line width=1.5pt, mark size=3.0pt, mark=otimes, mark options={solid, red}]
  table[row sep=crcr]{%
20	3.55\\
30	3.4\\
40	4.8\\
50	4.99\\
60	5.2\\
70	5.33\\
80	5.44\\
90	5.72\\
};

\end{axis}
\end{tikzpicture}%
}\subcaption{Average Path Length}\label{fig:apt}
	\end{minipage}\qquad
	\caption{Energy Efficiency and Performance of the MEPT heuristics according to the predicted $Umin$ and $Umax$ values for the Abilene, GEANT, and Nobel-Germany topology traces}
	\label{fig:energyandperformance}	
\end{figure}

Figure \ref{fig:ES} and \ref{fig:apt} illustrates the energy efficiency and average path length of the MEPT heuristics according to the predicted $Umin$ and $Umax$ parameters after applying the Refine heuristics. Results show that MEPT achieves an energy saving of 48\% for low traffic. The trends in energy saving are indirectly proportional to the traffic volume. This shows MEPT algorithm makes traffic proportional energy saving. The average path length measurement shows that increase in traffic increases the average path length. This is because as the volume of traffic increases, the shortest paths become overloaded, because of flows have to be re-routed to longer paths.

The significance of the machine learning method lies in the fact that it predicts $Umin$ and $Umax$ for a given new traffic. Getting the optimal values the parameters increases the EPT value. Maximum EPT value increases utilities of links, energy saving, and also maintains an acceptable network performance.

\section{Conclusion and Future Work} \label{sec:conclusion}

SDN is a very powerful networking paradigm that allows flexibility in the control and management of a network through re-routing flows in order to achieve efficiency, performance, load balancing, and security.  In this work, we proposed MER-SDN, a machine learning based framework for traffic aware energy efficient routing in SDN. We used topology and traffic as features to train our model. We also employed PCA and achieved a feature size reduction of more than 65\% on real-world network topology and dynamic traffic traces. Particularly, we tested MER-SDN to predict $Umin$ and $Umax$ parameters for the energy efficient MEPT heuristics algorithm. In addition to the prediction, we also proposed a heuristics to increase the accuracy of the predictor to 100\%. Experiment results show that the accuracy of predicting $Umin$ and $Umax$ are more than 70\%. The refining heuristics algorithm converges to the optimal $Umin$ and $Umax$ values 15 to 25 times faster than the brute force method.

Our framework is tested by taking snapshots of historical traffic traces. As future work, we plan to incorporate time dimension to the features. The effect of the traffic volume on the accuracy of our model would also be discussed. We aim at using reinforcement learning technique to help the controller to be self and incrementally learn and predict in a dynamic environment.  We also plan to add link and switch status as a feature, and conduct comprehensive experiments on the effect of the energy saving approach on throughput and delay. 

\bibliographystyle{IEEEtran} 
\bibliography{MLESDN}

\end{document}